# Scaling Invariance of Spatial Autocorrelation in Urban Built-Up Area

Meng Fu, Yanguang Chen*

(Department of Geography, College of Urban and Environmental Sciences, Peking University, Beijing 100871, China; chenyg@pku.edu.cn)

**Abstract:** City is proved to be a scale-free phenomenon, and spatial autocorrelation is often employed to analyze spatial redundancy of cities. Unfortunately, spatial analysis results deviated practical requirement in many cases due to fractal nature of cities. This paper is devoted to revealing the internal relationship between the scale dependence of Moran's I and fractal scaling. Mathematical reasoning and empirical analysis are employed to derive and test the model on the scale dependence of spatial autocorrelation. The data extraction way for fractal dimension estimation is box-counting method, and parameter estimation relies on the least squares regression. In light of the locality postulate of spatial correlation and the idea of multifractals, a power law model on Moran's I changing with measurement scale is derived from the principle of recursive subdivision of space. The power exponent is proved to be a function of fractal dimension. This suggests that the numerical relationship between Moran's I and fractal dimension can be established through the scaling process of granularity. An empirical analysis is made to testify the theoretical model. It can be concluded that spatial autocorrelation of urban built-up area has no characteristic scale in many cases, and urban spatial analysis need new thinking.

## 1 Introduction

Urban studies depends on spatial analysis, which needs robust measurements. However, scale-free property of cities affects the effect of spatial measurement. In fact, besides cities, the scale-free phenomena are very common in geographical world. What is called "scale-free" means has no characteristic scale, or the measurement results depend on measurement scales. Scientific research consists of description and understanding (Kane, 2005; Henry, 2002). Description relies on measure and mathematics (Henry, 2002). The application of mathematical methods in scientific research mainly plays two functions: one is to establish models, and the other is to sort out observation data. In any case, the basic premise of mathematical modeling and quantitative analysis is to find characteristic scale (Takayasu, 1990). However, it is often difficult to find effective characteristic scales in complex systems such as cities. In fact, scale dependence can be divided into two cases. First, the relationship between scale and measure can be described by functions with characteristic scales, such as exponential function or logarithmic function. Such scale dependence phenomena are not complex, and the characteristic scales can be found based on function relations. Second, the relationship between scale and measure must be described by power law, in which no characteristic scale can be found. Such scale dependence phenomena represent complex system, in which characteristic scale cannot be found and conventional quantitative analysis method fails. Cities belong to the second type of scale dependence phenomena, i.e., scale-free phenomena. Scaling analysis is a new way of exploring scale-free problems. The relationship between scale dependence and scaling has been concerned for a long time (Su et al., 2001; Wu, 1996). One of the effective tools for scaling analysis is fractal geometry (Mandelbrot, 1982).

Spatial autocorrelation is one of the basic methods for geospatial analysis. The important tool for spatial statistics has been applied to urban studies. One of the basic measures of spatial autocorrelation analysis is Moran's *I*, which is proved to be the eigenvalue of generalized spatial weights matrix (Chen, 2013). In this sense, Moran's *I* is a statistic representing characteristic scale. The premise of validity of Moran's *I*, as a characteristic scale, is its scale independence. In other words, it has robustness of measurement. However, a large number of studies have shown that the calculation results of Moran's *I* of urban space bear scale dependence. On the one hand, granularity of spatial measurement sometimes affects calculation results significantly (e.g., Chou, 1991; Feng et al., 2016; Overmars et al., 2003; Qi & Wu, 1996); on the other hand, threshold distance defining spatial contiguity matrix also affects the final value of Moran's *I* significantly (e.g., Bjørnstad & Falck, 2001; De Knegt et al., 2010; Getis & Ord, 1992; Legendre & Legendre, 1998; Odland, 1988; Ord & Getis, 1995). If an urban model has characteristic scales, Moran's *I* can be regarded as a characteristic parameter; if the model has no characteristic scale, it relates to scaling process. There are two ways to solve the scaling problem. One is to construct spatial autocorrelation function based on variable scales, so as to replace the spatial autocorrelation index; the other is to reveal the internal relationship between scale dependence and fractal scaling, and convert the spatial autocorrelation index into scaling index (Chen, 2021).

Fractals suggest optimized structure of a self-organized system, while spatial autocorrelation implies information redundancy in the system. The relationship between fractal scaling and spatial autocorrelation helps to understand the spatial complexity of cities and dynamic mechanisms of urban evolution. This paper is devoted to revealing the underlying scaling properties and fractal features of spatial autocorrelation of urban form. The rest of this paper is arranged as follows: in Section 2, based on urban form, the mathematical model of scale dependence of Moran's $I$ is derived; the corresponding scaling exponent is proved to be a function of fractal dimension, and case analysis based on 19 largest Chinese cities is carried out. In Section 3, the results of case study are analyzed and the mathematical model derived before is verified; in Section 4, achievements, novelties and limitations of research methods in this paper are discussed; finally, In Section 5, main conclusions are drawn based on the research results and question discussion.

## 2 Materials and methods

### 2.1 Moran's $I$ and correlation dimension

Before theoretical derivation, two postulates are given in this paper. First, the spatial pattern of urban built-up area is of multifractal nature. Second, the spatial correlation of different areas is local. There are a lot of research results of multifractal urban spatial pattern, which is not a strict postulate. The postulate of locality restricts construction of spatial contiguity matrix in this paper. Moran's $I$ is the basis of theoretical derivation and tool for empirical analysis in this paper, so it is necessary to show it in advance. The calculation formula of Moran's $I$ can be expressed as

$$I = \frac{N \sum_{i=1}^{N} \sum_{j=1}^{N} v_{ij}(x_i - \bar{x})(x_j - \bar{x})}{\sum_{i=1}^{N} \sum_{j=1}^{N} v_{ij} \cdot \sum_{i=1}^{N} (x_i - \bar{x})^2} = \frac{\sum_{i=1}^{N} \sum_{j=1}^{N} v_{ij}(x_i - \bar{x})(x_j - \bar{x})}{V \sum_{i=1}^{N} \sum_{j=1}^{N} v_{ij}}, \quad (1)$$

where

$$V = \frac{1}{N} \sum_{i=1}^{N} (x_i - \bar{x})^2, \quad (2)$$

is the population variance of variable $x$, $v_{ij}$ is the contiguity measure, which can be measured by the reciprocal of distance (numerical variable) or the adjacency relationship (virtual variables 0 and 1) between $i$ and $j$. Scale effects in spatial analysis include both extent and granularity. In previous research on the scale dependence of Moran's *I*, a significant focus has been placed on granularity (Chou, 1991). There are two main ways to achieve different granularities in existing literature: One is segmentation method, which is to divide study area with several grids first, then divide one grid into two, two into four......, the more times of segmentation, the smaller the granularity (e.g., Chou, 1991). The other is aggregation method, which is to aggregate pixels at different levels based on the original spatial pattern of study area, the higher the aggregation level, the larger the granularity (e.g., Qi & Wu, 1996; Overmars et al., 2003; Feng et al., 2016). The segmentation method operates from top to bottom, which is similar to the calculation process of box dimension, while there are few data points. The aggregation method operates from bottom to top, which is similar to renormalization. There are more data points as the granularities are usually arithmetic series, while the total area covered by pixels is actually not equal as the granularities are not multiple.

In order to help readers understand the derivation process of equations proposed in this paper, it is necessary to explain data processing method in advance. Based on the advantages and disadvantages of the above two methods, this paper selects the segmentation method to extract granularities, and makes slight adjustments with reference to the box-counting method: first, select the smallest circumscribed rectangle in study area as the first-level box, then divide the box into four, four into sixteen......, where the granularities are multiple. Specifically, assuming the study area is divided into $N$ non-empty boxes in granularity $\varepsilon$, then $x_i$ and $x_j$ in Eq. 1 are the values of variable $x$ in the $i$th and $j$th box respectively, $\bar{x}$ are the average values of $x$. Referring to Chou (1991), variable $x$ in this paper is defined in two ways to compare the influence of variable types on the scale dependence of Moran's *I*. One type of variable $x$ is numerical variable, where $x$ is the proportion of built-up area in each box to box area, that is, $x_i = A_i /\varepsilon^2$ ($A_i$ is the area of built-up area in box $i$). The variable is called p variable. The other type is classification variable, $x$ takes 0 or 1 indicating whether built-up area is dominant in each box, that is, whether its area exceeds 50%. If it is, $x$ takes 1, otherwise, $x$ takes 0. The variable is called t variable. The classification variable t is equivalent to raster images with the same resolution. It helps reduce the fragmentation of the original spatial pattern, similar to how remote sensing images work (Zuo, 2011). However, it is affected by mixed pixels and may not accurately represent the original spatial pattern. Additionally, increasing granularity through the use of t variables can further distort spatial patterns and produce results similar to the process of aggregating pixels by majority in remote sensing images. On the other hand, the numerical variable p is closer to the original spatial pattern. In addition, $v_{ij}$ in Eq. 1 represents the value of contiguity matrix **V** at the corresponding position. There are at least four types of contiguity matrix (Chen, 2012). Based on the postulate of absolute locality in this paper, the simplest rook contiguity matrix is selected for **V**, that is, when the *i*th and *j*th box have common edges, $v_{ij}$ takes 1, otherwise $v_{ij}$ takes 0.

As a preparation for theoretical derivation, the definition of correlation dimension is explained. The scale dependence of spatial measurement is essentially a scaling phenomenon. Fractal geometry is a powerful tool for scaling analysis, and fractal dimension is the basic parameter of fractal system (Batty & Longley, 1994; Frankhauser, 1998; Mandelbrot, 1982). There are three commonly used parameters in multifractal spectrum: capacity dimension, information dimension and correlation dimension (Grassberger, 1983). The most widely used measurement method of fractal dimension is box-counting method, among which the most convenient one is functional box method (Lovejoy et al., 1987; Chen, 1995). The specific calculation steps are as follows: first, select the smallest circumscribed rectangle of study area as the first-level box; then divide the box into four, four into sixteen......, measure the box size $\varepsilon$ in each level and the proportion $p_i$ of the area of built-up area in the *i*th box to that in the total area, based on which capacity dimension and correlation dimension can be calculated. Based on the postulate of uniform distribution, the probability $p_i$ is measured by virtual variables 0 or 1, then the definition of capacity dimension can be derived. The calculation formula of capacity dimension is

$$D_0 = -\lim_{\varepsilon \to 0} \frac{\ln N(\varepsilon)}{\ln \varepsilon} = -\lim_{\varepsilon \to 0} \frac{\ln \sum_{i}^{N(\varepsilon)} p_i^0}{\ln \varepsilon}, \tag{3}$$

where $D_0$ is the capacity dimension in the study area, $N(\varepsilon)$ is the number of non-empty boxes under scale $\varepsilon$. Eq. 3 is equivalent to the power law relation defining the box dimension

$$N(\varepsilon) = \eta \varepsilon^{-D_0}, \tag{4}$$

where $\eta$ is the proportional coefficient, $\eta = 1$ theoretically.

Correlation dimension is defined based on the second order moment of probability, which is equivalent to the second-order Renyi entropy. The calculation formula of correlation dimension is

$$D_2 = \lim_{\varepsilon \to 0} \frac{\ln C(\varepsilon)}{\ln \varepsilon} = \lim_{\varepsilon \to 0} \frac{\ln \sum_{i}^{N(\varepsilon)} p_i^2}{\ln \varepsilon}, \tag{5}$$

where $D_2$ is the correlation dimension, $C(\varepsilon)$ is the correlation function under scale $\varepsilon$. Based on step function, the correlation function can be simplified as (Grassberger & Procassia, 1983)

$$C(\varepsilon) = \frac{1}{N(\varepsilon)^2} \sum_{i}^{N(\varepsilon)} \sum_{j}^{N(\varepsilon)} \theta(\varepsilon - d_{ij}) \propto \sum_{i}^{N(\varepsilon)} p_i^2, \qquad (6)$$

where $\theta$ is the Heaviside function

$$\theta(\varepsilon - d_{ij}) = \begin{cases} 1, & d_{ij} \leq \varepsilon \\ 0, & d_{ij} > \varepsilon \end{cases}. \qquad (7)$$

Capacity dimension $D_0$ and correlation dimension $D_2$ represent the degree of spatial filling and spatial correlation respectively, and $D_0 > D_2$. It can be derived from Eqs. 5 and 6 that

$$C(\varepsilon) = \sum_{i=1}^{N(\varepsilon)} p_i^2 = K\varepsilon^{D_2}, \qquad (8)$$

where the proportional coefficient $K = 1$ in standard case.

## 2.2 Derivation of scaling relation

Based on the preparation of above concepts, methods and formulas, the relationship between Moran's $I$ and fractal dimension can be formally derived. With the results of this derivation, the scale dependence of Moran's $I$ of cities can be better understood. Moran's $I$ is essentially a generalized correlation function, which can be related to correlation dimension (Chen, 2021). On the one hand, for numerical variable p, $x_i$ in Eq. 1 can be associated with probability measure $p_i$, that is

$$x_i = \frac{A_i}{\varepsilon^2} = \frac{A_i}{A} \cdot \frac{A}{\varepsilon^2} = p_i \cdot \frac{A}{\varepsilon^2}, \qquad (9)$$

where $A$ is the total area of built-up area of a city. Then there is

$$\bar{x} = \frac{A}{\varepsilon^2 N(\varepsilon)} \cdot \sum_{i}^{N(\varepsilon)} p_i = \frac{A}{\varepsilon^2 N(\varepsilon)}. \qquad (10)$$

Here we use the normalization property of probability, that is, the sum of probabilities is 1. It is worth noting that for fractal phenomena, the total area $A$ is a variable in theory; but in practice, the spatial pattern under certain resolution is actually prefractal rather than real fractal, so $A$ can be treated as a constant.

The calculation result of Moran's $I$ depends on the definition of spatial weights matrix, which depends on the definition of spatial contiguity. Based on localized contiguity represented by 0 and 1, we can not only calculate a kind of Moran's $I$, but also define spatial correlation function. This means that with the contiguity relationship represented by virtual variable, Moran's *I* and correlation function $C(\varepsilon)$ can be linked. Specifically, based on the contiguity matrix **V** represented by variable 0 and 1, Eqs. 1 and 6 can be linked. In fact, under box size $\varepsilon$, it can be approximated that only the boxes adjacent to box *i* are less than $2\varepsilon$ away from it, so $\theta(2\varepsilon-d_{ij}) = 1$; the other boxes are more than $2\varepsilon$ away from box *i*, so $\theta(2\varepsilon-d_{ij}) = 0$. In this case, when **V** takes rook contiguity matrix, only the $\theta$ function of box *i* and the four boxes adjacent to and on the top, bottom, left and right of it take 1, the rest take 0; when **V** takes queen contiguity matrix, the $\theta$ function of box *i* and the eight boxes adjacent to and on the top, bottom, left, right, top left, top right, lower left and lower right of it take 1, the rest take 0. In this paper, **V** takes rook contiguity matrix, so Eq. 6 can be rewritten as

$$C(2\varepsilon) = \sum_{i}^{N(\varepsilon)} p_i(p_i + p_{it} + p_{ib} + p_{il} + p_{ir}) = C(\varepsilon) + \sum_{i}^{N(\varepsilon)} \sum_{j}^{N(\varepsilon)} v_{ij} p_i p_j, \quad (11)$$

where the subscripts *it*, *ib*, *il* and *ir* represent the boxes adjacent to and on the top, bottom, left and right of box *i* respectively. Based on the postulate of scaling property of spatial relationship, introduce Eq. 8 into Eq. 11 to obtain

$$\sum_{i=1}^{N(\varepsilon)} \sum_{j=1}^{N(\varepsilon)} v_{ij} p_i p_j = (2^{D_2} - 1)C(\varepsilon), \quad (12)$$

further, introducing Eqs. 9, 10 and 12 into Eq. 1 yields

$$I(\varepsilon) = \frac{\frac{A^2}{\varepsilon^4} \sum_{i=1}^{N(\varepsilon)} \sum_{j=1}^{N(\varepsilon)} [v_{ij} p_i p_j - \frac{v_{ij} p_i}{N(\varepsilon)} - \frac{v_{ij} p_j}{N(\varepsilon)} + \frac{v_{ij}}{N(\varepsilon)^2}]}{\sum_{i=1}^{N(\varepsilon)} \sum_{j=1}^{N(\varepsilon)} v_{ij} \cdot V(\varepsilon)} = \frac{A^2[(2^{D_2} - 1)C(\varepsilon) - \frac{2\sum_{i=1}^{N(\varepsilon)} p_i \sum_{j=1}^{N(\varepsilon)} v_{ij}}{N(\varepsilon)} + \frac{\sum_{i=1}^{N(\varepsilon)} \sum_{j=1}^{N(\varepsilon)} v_{ij}}{N(\varepsilon)^2}]}{\sum_{i=1}^{N(\varepsilon)} \sum_{j=1}^{N(\varepsilon)} v_{ij} \cdot \varepsilon^4 V(\varepsilon)}.$$

$$(13)$$

According to Eq. 2, $V(\varepsilon)$ is the population variance of variable *x* under granularity $\varepsilon$.

A skill of mathematical reasoning is to approximate with the help of limit conditions, and approximate results usually do not affect the effect of empirical analysis. In this paper **V** takes rook contiguity matrix, then when $\varepsilon \to 0$ that is $N(\varepsilon) \to \infty$, the sum of $v_{ij}$ by column is close to 4, so Eq. 13 is simplified as

$$I(\varepsilon) \cong \frac{A^2[(2^{D_2} - 1)C(\varepsilon) - 4/N(\varepsilon)]}{4\varepsilon^4 N(\varepsilon) V(\varepsilon)} = \frac{A^2[(2^{D_2} - 1)N(\varepsilon)C(\varepsilon) - 4]}{4\varepsilon^4 N(\varepsilon)^2 V(\varepsilon)}, \quad (14)$$

it can be seen from Eqs. 4 and 8 that

$$N(\varepsilon)C(\varepsilon) = \eta K \varepsilon^{D_2 - D_0}, \tag{15}$$

where $D_0 > D_2$. Then when $\varepsilon \to 0$, there is $N(\varepsilon)C(\varepsilon) \to \infty$, so

$$I(\varepsilon) \cong \frac{(2^{D_2}-1)A^2 N(\varepsilon)C(\varepsilon)}{4\varepsilon^4 N(\varepsilon)^2 V(\varepsilon)} = \frac{(2^{D_2}-1)A^2 C(\varepsilon)}{4\varepsilon^4 N(\varepsilon)V(\varepsilon)} = \lambda \frac{C(\varepsilon)}{\varepsilon^4 N(\varepsilon)V(\varepsilon)}, \tag{16}$$

where

$$\lambda = \frac{1}{4}(2^{D_2}-1)A^2, \tag{17}$$

is a proportional constant. Then Eq. 16 becomes

$$I(\varepsilon)V(\varepsilon) = \lambda \frac{C(\varepsilon)}{\varepsilon^4 N(\varepsilon)} = \lambda' \varepsilon^{D_0 + D_2 - 4}, \tag{18}$$

where $\lambda' = \lambda K/\eta$. It can be seen that when $V(\varepsilon)$ shows a power law against $\varepsilon$, $I(\varepsilon)$ follows a power law against $\varepsilon$ as well.

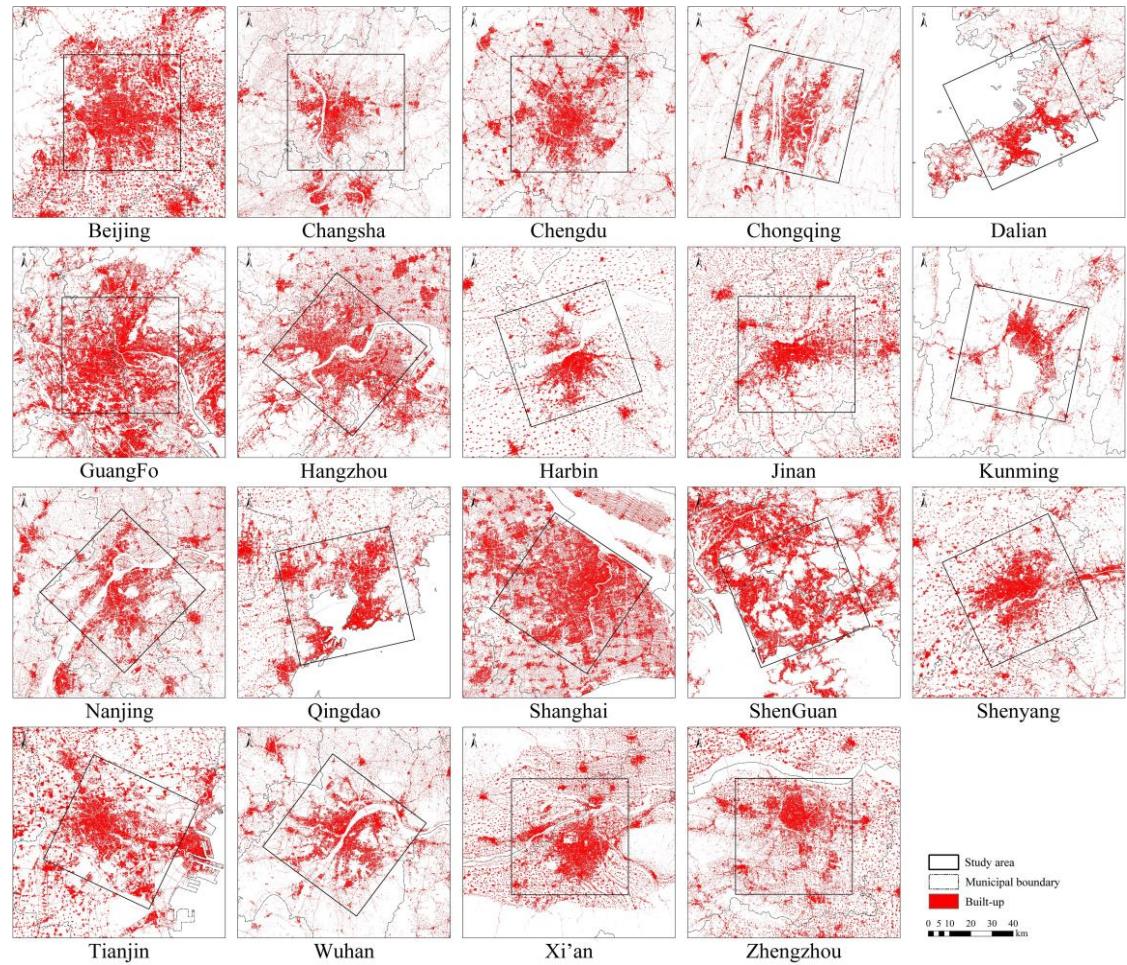

**Fig. 1** 19 research objects and corresponding research scopes. Contains modified Copernicus Sentinel data (2021) processed by ESA WorldCover consortium.

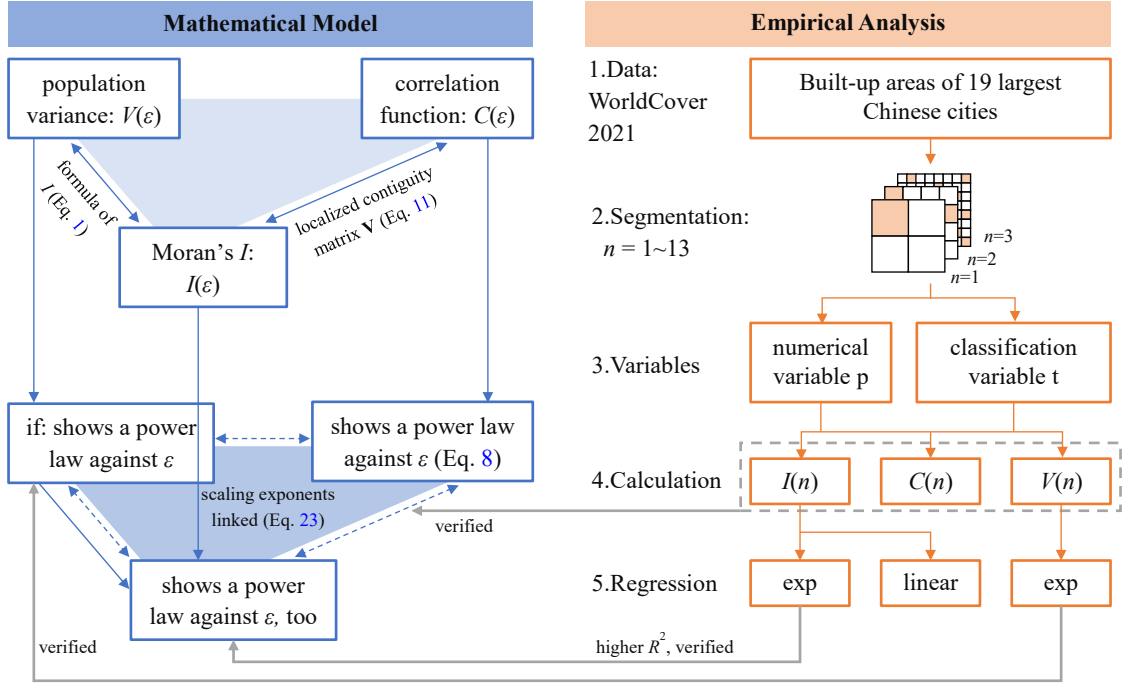

**Fig. 2** Workflow of this paper. Moran's *I* can be linked to correlation function by contiguity matrix, so the scaling exponent that Moran's *I* changing with granularity can be linked with correlation dimension. Taking 19 largest Chinese cities as examples, the derivation above is verified.

## 2.3 Study area and data processing

Success of modern science lies in its great emphasis on the role of quantifiable data and its interaction with models. Both data and models are necessary and indispensable for the advancement of human understanding of problems: on the one hand, data generates formalized facts and imposes constraints on models; on the other hand, models are significant to understand the operation principle and action process of system (Louf & Barthelemy, 2014). The above relationship between data and models affects all scientific fields, and this research is no exception. The empirical analysis in this paper focuses on the urban areas of 21 largest cities in China to validate previous derivation results. The cities include 7 megacities with a permanent urban population exceeding 10 million: Shanghai, Beijing, Shenzhen, Chongqing, Guangzhou, Chengdu and Tianjin; as well as 14 supercities with a permanent urban population ranging from 5 to 10 million: Wuhan, Dongguan, Xi'an, Hangzhou, Foshan, Nanjing, Shenyang, Qingdao, Jinan, Changsha, Harbin, Zhengzhou, Kunming and Dalian. Since the built-up areas of Guangzhou and Foshan, as well as Shenzhen and Dongguan, have already interconnected, this paper combines them into single cities respectively, referred to as GuangFo and ShenGuan. So there are a total of 19 research objects (Fig. 1). The research scope covers a 55km square area around the central part of each city, which is sufficient to cover the most urbanized regions in all cities. WorldCover dataset of 2021, a land cover dataset published by the European Space Agency with a resolution of 10m and an overall accuracy of 76.7%, is selected as data source (Zanaga et al., 2022). This paper only selects one land cover class, built-up, among 11 land cover classes in this dataset, to represent urbanized area. Furthermore, the original images were processed using ArcGIS 10.3, and Moran's *I* was calculated using Python 3.8. Finally, Moran's *I* are calculated for segmentation times from 1 to 13, corresponding to granularities of about 27.5km to 7m. The workflow of the study is illustrated in Fig. 2.

## 3 Results

After calculating Moran's *I* of urban form, we can examine the relationships between spatial autocorrelation measure and fractal scaling of cities. For the 19 largest Chinese cities listed in Fig. 1, Moran's *I* of built-up area in each city under different granularities are calculated respectively. For convenience, we use segmentation time *n* to replace granularity *ε* as abscissa in the following plots, the results are shown in Fig. 3.

First, Moran's *I* show an upward trend as granularity becomes smaller. Except a few Moran's *I* that are negative under coarse granularities, the rest are all positive, indicating that the spatial pattern of built-up areas is positively correlated. Moran's *I* fluctuate greatly under coarse granularities, but their trend becomes clear as the increase of segmentation time *n*, that is, Moran's *I* increases with *n* after 6~7 times of segmentation. Compared with t variables, the fluctuation of p variables is smaller and their trend is more obvious. The above trend is basically consistent with existing research results (Chou, 1991; Qi & Wu, 1996; Bu et al., 2003; Xu et al., 2004; Tan et al., 2005; Xu et al., 2007).

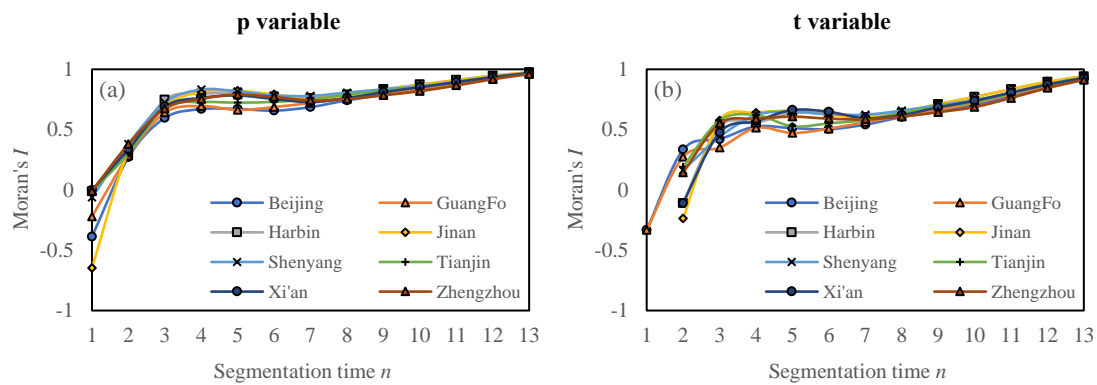

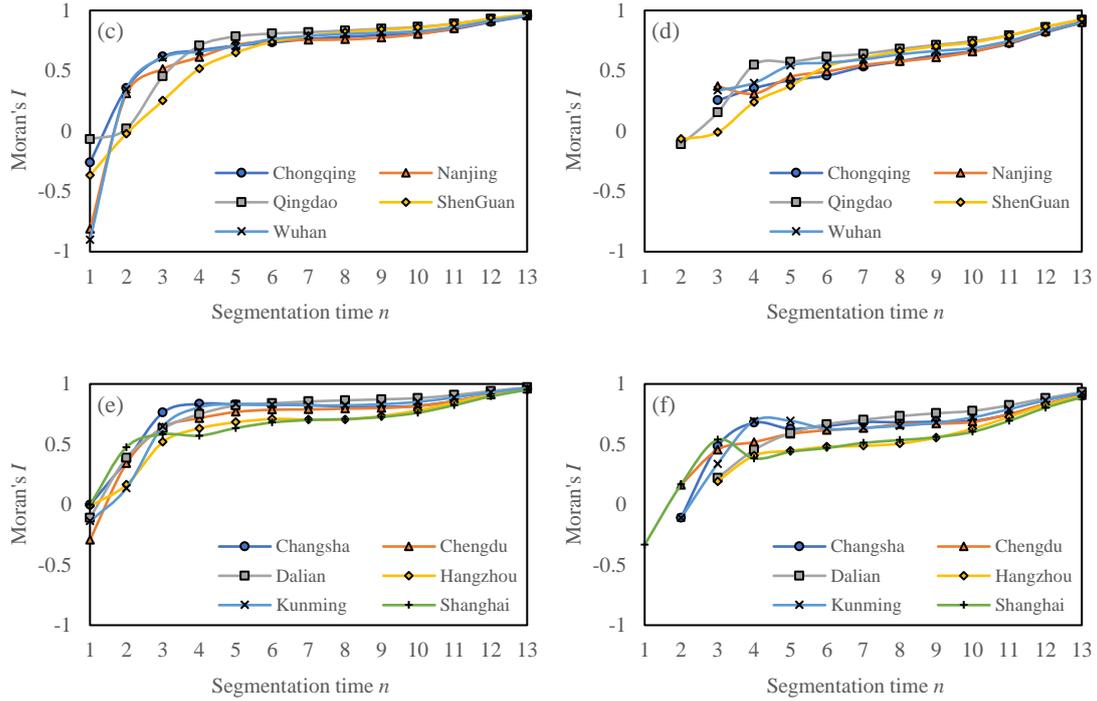

**Fig. 3** The change of Moran's *I* with segmentation time *n* in 19 largest Chinese cities. All Moran's *I* show an upward trend with the increase of *n*. Moran's *I* of p variable (**a**, **c, e**) is larger and more stable than that of the corresponding t variable (**b, d, f**). The 19 cities can be classified into three groups according to how their turning points for p and t variables are distributed.

Second, the trends of Moran's *I* changing with granularity of two types of variables (p, t) are similar yet diversified. On the whole, Moran's *I* of p variable (Fig. 3a, c, e) is larger than that of the corresponding t variable (Fig. 3b, d, f), and the turning points where Moran's *I* change from fluctuation to trend with the increase of *n* are different. But with the increase of segmentation time *n*, Moran's *I* of t variable increases faster so the gap between the two is narrowed. Although the overall trends of the two variables are not the same, they tend to be consistent after the turning point. This is because compared with p variable, t variable only takes the dominant land use class and loses the other, so the spatial autocorrelation is weakened and Moran's *I* is smaller. With the increase of granularity, the influence of mixed pixels of t variable increases, and the gap of Moran's *I* between the two variables also increases.

Last but not least, the trends of Moran's *I* changing with granularity vary across different cities as well. The turning points for the 19 cities vary, and they can be classified into three types based on their respective turning points for p and t variables (Table 1). (1) The first type of cities have a consistent changing trend of Moran's *I* between p and t variables, with turning points occurring early at $n$ = 5~7. These cities include Beijing, GuangFo, Harbin, Jinan, Shenyang, Tianjin, Xi'an, and Zhengzhou. These cities exhibit strong spatial heterogeneity at coarser granularities, leading to a smaller Moran's *I*. However, as granularity decreases, the spatial heterogeneity weakens, and Moran's *I* increases significantly. Besides, there is little difference between the spatial pattern of high-density built-up area (t variable) and overall built-up area (p variable). This indicates these cities have a clear polycentric structure with developed multifractal nature (Fig. 3a, b). (2) The second type of cities have an inconsistent changing trend of Moran's *I* between p and t variables, with turning points of t variables occurring early at $n$ = 5~7, while Moran's *I* of p variables stabilizes at high values. As granularity further decreases, Moran's *I* of p variables show an upward trend, and their turning points occur at $n$ = 8~10. This type of cities includes Chongqing, Nanjing, Qingdao, Shenzhen, and Wuhan. These cities exhibit relatively weak spatial heterogeneity at coarser granularities, leading to a higher value of Moran's *I* for p variables. However, if only high-density built-up area are considered (t variable), the spatial heterogeneity will significantly increase, resulting in a much smaller value of Moran's *I* for t variables than that for p variables. This indicates that these cities have small subsidiary-centers with developing multifractal nature (Fig. 3c, d). (3) The third type of cities have a consistent changing trend of Moran's *I* between p and t variables as well, but with turning points occurring later at $n$ = 8~10. This type of cities includes Changsha, Chengdu, Dalian, Hangzhou, Kunming, and Shanghai. These cities exhibit relatively weak spatial heterogeneity at coarser granularities, so Moran's *I* stabilize at high values. However, as granularity further decreases, Moran's *I* start to increase significantly. Furthermore, there is little difference between the spatial pattern of high-density built-up area (t variable) and overall built-up area (p variable). This indicates that these cities have a clear monocentric structure with undeveloped multifractal nature (Fig. 3e, f). Finally, as $n$ increases, Moran's *I* with smaller values generally increase faster, which results in a decrease in the gap of Moran's *I* in different cities. However, after the turning points, their relative size tends to be stable with little change.

**Table 1** Classification of 19 cities studied in this paper based on the changing trend of Moran's *I*.

| Type | Changing trend of Moran's *I* | Cities |
|---|---|---|
| 1. Polycentric structure, developed multifractal nature | Consistent changing trend between p and t variables; turning points occurred early at $n$ = 5~7. | Beijing, GuangFo, Harbin, Jinan, Shenyang, Tianjin, Xi'an, Zhengzhou |
| 2. Small subsidiary- | Inconsistent changing trend between p and t variables; | Chongqing, Nanjing, |
| centers, developing multifractal nature | turning points of t variables occurred early at $n = 5\sim7$, while that of p variables occurred later at $n = 8\sim10$. | Qingdao, Shenzhen, Wuhan |
| 3. Monocentric structure, undeveloped multifractal nature | Consistent changing trend between p and t variables; turning points occurred later at $n = 8\sim10$. | Changsha, Chengdu, Dalian, Hangzhou, Kunming, Shanghai |

Regression analysis is carried out on the scattered points then. Based on the overall trend, the change of Moran's *I* with segmentation time *n* is speculated to follow a linear or an exponential function. The model of linear function is

$$I(n) = a + bn, \qquad (19)$$

where *n* is segmentation time, *I(n)* is the corresponding Moran's *I*, *a* is the constant term, and *b* is the coefficient. The model of exponential function is

$$I(n) = \alpha \beta^n, \qquad (20)$$

where *α* is the coefficient, *β* is the base number, and other symbols have the same meaning as Eq. 19. Since *ε* shows an exponential law against *n* that is $\varepsilon \propto 2^{-n}$, the linear or exponential relationship between Moran's *I* and *n* in Eqs. 19 and 20 can be transformed into a logarithmic or power relationship between Moran's *I* and *ε*:

$$I(\varepsilon) = a' - b' \ln \varepsilon, \qquad (21)$$

$$I(\varepsilon) = \alpha' \varepsilon^{\beta'}, \qquad (22)$$

where *ε* is granularity, *I(ε)* is the corresponding Moran's *I*, $b' = b/\ln 2$, $\beta' = -\log_2 \beta$. Eq. 21 is a logarithmic function, which is consistent with the research results of Chou (1991) and Zhang et al. (2019), so the linear function in Eq. 19 is empirically supported. Eq. 22 is a power function, which is consistent with the calculation formula of box dimension. In fact, according to Chou's (1991) explanation for the scale dependence of Moran's *I*, as the increase of *n*, the homogeneous pixels increase in 2 dimensions while the heterogeneous ones increase in 1 dimension, it is speculated that the spatial autocorrelation pattern of built-up area changes between 1 and 2 dimensions with granularity. That is, the change of Moran's *I* may be fractal and in the form of Eq. 22, so the exponential function in Eq. 20 is theoretically supported. Eqs. 19 and 20 were used to fit the data after turning points in Fig. 3, and the results are shown in Table 2.

**Table 2** Fitting results of Moran's *I* changing with segmentation time *n*. Moran's *I* have a higher goodness of fit to Eq. 20 except for the part marked red.

| City | Based on numerical variable p | | | Based on classification variable t | | |
|---|---|---|---|---|---|---|
| | Turning point | $R^2$ for Eq. 19 | $R^2$ for Eq. 20 | Turning point | $R^2$ for Eq. 19 | $R^2$ for Eq. 20 |
| Beijing | 6 | 0.9980 | **0.9959** | 6 | 0.9915 | 0.9984 |
| Changsha | 10 | 0.9957 | 0.9969 | 10 | 0.9942 | 0.9965 |
| Chengdu | 10 | 0.9975 | 0.9979 | 10 | 0.9972 | 0.9975 |
| Chongqing | 10 | 0.9934 | 0.9955 | 5 | 0.9840 | 0.9923 |
| Dalian | 9 | 0.9706 | 0.9740 | 9 | 0.9763 | 0.9816 |
| GuangFo | 5 | 0.9965 | 0.9987 | 5 | 0.9929 | 0.9990 |
| Hangzhou | 8 | 0.9869 | 0.9913 | 8 | 0.9899 | 0.9951 |
| Harbin | 7 | 0.9970 | 0.9935 | 7 | 0.9959 | 0.9917 |
| Jinan | 7 | 0.9978 | 0.9959 | 7 | 0.9973 | 0.9944 |
| Kunming | 9 | 0.9873 | 0.9901 | 9 | 0.9939 | 0.9957 |
| Nanjing | 9 | 0.9888 | 0.9923 | 5 | 0.9664 | 0.9888 |
| Qingdao | 9 | 0.9742 | 0.9786 | 5 | 0.9777 | 0.9917 |
| Shanghai | 9 | 0.9891 | 0.9923 | 9 | 0.9836 | 0.9907 |
| ShenGuan | 8 | 0.9737 | 0.9797 | 7 | 0.9875 | 0.9938 |
| Shenyang | 7 | 0.9926 | 0.9963 | 7 | 0.9871 | 0.9954 |
| Tianjin | 6 | 0.9941 | 0.9973 | 6 | 0.9897 | 0.9966 |
| Wuhan | 10 | 0.9949 | 0.9962 | 5 | 0.9481 | 0.9743 |
| Xi'an | 7 | 0.9980 | 0.9952 | 7 | 0.9921 | 0.9952 |
| Zhengzhou | 7 | 0.9733 | 0.9812 | 7 | 0.9617 | 0.9783 |

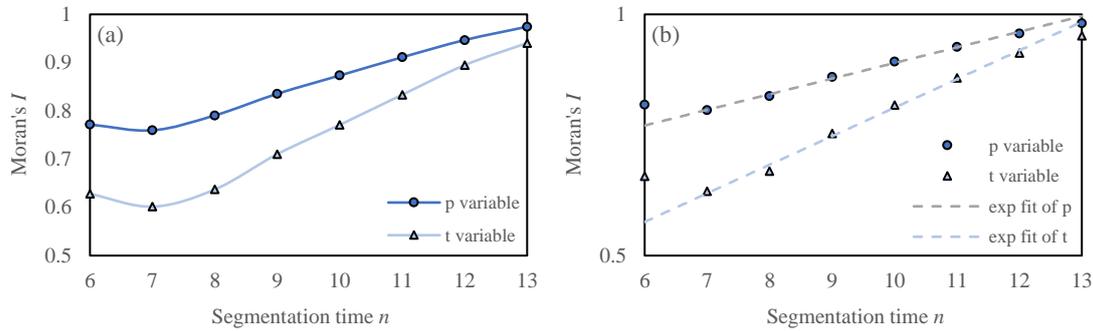

**Fig. 4** The logarithmic change of Moran's $I$ with segmentation time $n$, taking Harbin as example. In linear scale, the curves after the turning points all accelerated first at $n = 7\sim12$ and then stabilized at $n = 13$ (**a**); in logarithmic scale, the curves after the turning points all increase linearly first at $n = 7\sim12$ and then tend to be flat at $n = 13$ (**b**).

It can be seen that, although the turning points are not the same, after that the change of Moran's $I$ of most variables with $n$ are more exponential, that is, Moran's $I$ change with the power of $\varepsilon$. However, Moran's $I$ of six variables, Beijing_p, Xi'an_p, Harbin_p, Harbin_t, Jinan_p and Jinan_t, show a higher goodness of fit to linear function, that is, Moran's $I$ change logarithmically with $\varepsilon$, which is consistent with the research results of Chou (1991) and Zhang et al. (2019). The six curves after the turning points all accelerated first and then stabilized (Fig. 4a); taking the logarithm of Moran's $I$, the six curves after the turning points all increase linearly first and then tend to be flat (Fig. 4b). This is because Moran's $I$ has an upper limit of 1, and the Moran's $I$ of these six variables are close to or even more than 0.95 when $n = 13$. So Moran's $I$ is significantly squeezed by the upper limit, its growth rate slows down, and its change with $\varepsilon$ is more logarithmic.

However, it is not sufficient to determine the mathematical model for scale dependence of Moran's $I$ simply relying on $R^2$, further analysis based on model parameters is required. When the observed data follow Eq. 19, the scale dependence has a characteristic scale, and the parameter $b'$ in Eq. 21 is the eigenvalue of Moran's $I$. On the other hand, when the observed data follow Eq. 20, the scale dependence has no characteristic scale, and scaling analysis should be carried out based on the parameter $\beta'$ in Eq. 22. Moreover, if $b'$ exceeds the range of Moran's $I$, indicating that the center of Moran's $I$ is outside the range of its values, then the scale dependence also has no characteristic scale. Table 3 compares the range of Moran's $I$ and the characteristic scale $b'$ in each city. It is found that in all cities, $b'$ exceeds the range of Moran's $I$. Therefore, despite having a high $R^2$ for Eq. 19, the scale dependence of Moran's $I$ is essentially scale-free, and can be approximated by a power law, namely Eq. 22. In fact, Zhang et al. (2019) also found that Moran's $I$ follows a logarithmic relationship with resolution, but the eigenvalue of its logarithmic function also exceeded the range of Moran's $I$, further supporting the idea that the scale dependence of Moran's $I$ is scale-free.

**Table 3** The range of Moran's $I$ and the characteristic scale $b'$ of logarithmic fit in each city. $b'$ in all cities exceeds the range of Moran's $I$, indicating the scale dependence of Moran's $I$ is essentially scale-free and can be approximated by a power law.

| City | p variable | | t variable | |
|---|---|---|---|---|
| | $b'$ | Range of $I$ | $b'$ | Range of $I$ |
| Beijing | 0.149 | [0.657, 0.964] | 0.201 | [0.503, 0.921] |
| Changsha | 0.058 | [0.828, 0.957] | 0.107 | [0.652, 0.904] |
| Chengdu | 0.080 | [0.786, 0.958] | 0.130 | [0.617, 0.906] |
| Chongqing | 0.097 | [0.733, 0.954] | 0.196 | [0.460, 0.897] |
| Dalian | 0.059 | [0.839, 0.971] | 0.120 | [0.665, 0.932] |
| GuangFo | 0.134 | [0.687, 0.966] | 0.198 | [0.508, 0.923] |
| Hangzhou | 0.127 | [0.709, 0.957] | 0.213 | [0.479, 0.901] |
| Harbin | 0.109 | [0.772, 0.974] | 0.170 | [0.628, 0.941] |
| Jinan | 0.103 | [0.791, 0.974] | 0.171 | [0.623, 0.943] |
| Kunming | 0.067 | [0.830, 0.966] | 0.145 | [0.627, 0.922] |
| Nanjing | 0.101 | [0.747, 0.957] | 0.188 | [0.493, 0.900] |
| Qingdao | 0.073 | [0.808, 0.965] | 0.143 | [0.616, 0.922] |
| Shanghai | 0.129 | [0.680, 0.951] | 0.195 | [0.469, 0.890] |
| ShenGuan | 0.100 | [0.741, 0.967] | 0.175 | [0.536, 0.924] |

| | | | | | |
|---|---|---|---|---|---|
| Shenyang | 0.095 | [0.783, 0.970] | 0.153 | [0.627, 0.930] |
| Tianjin | 0.116 | [0.730, 0.969] | 0.183 | [0.552, 0.929] |
| Wuhan | 0.086 | [0.762, 0.957] | 0.154 | [0.566, 0.902] |
| Xi'an | 0.117 | [0.754, 0.968] | 0.157 | [0.648, 0.928] |
| Zhengzhou | 0.102 | [0.770, 0.960] | 0.160 | [0.591, 0.911] |

In summary, the change of Moran's *I* with granularity follows a power function, but there is a scaling range: when granularity is too coarse, Moran's *I* fluctuates greatly with large noise; when granularity is too fine, Moran's *I* is squeezed by the upper limit of 1, and its change slows down; when granularity is moderate, it is the scaling range where Moran's *I* follows a power law against granularity, corresponding to segmentation time $n$ of about 6~13 and granularity $\varepsilon$ of about 1000m~10m. And the logarithmic change of Moran's *I* with granularity is actually the latter part of the whole changing interval, corresponding to the scaling range and the part with finer granularity. However, the high Goodness of fit for logarithmic function is a false result, the characteristic scale of the equation exceeds the range of Moran's *I*, indicating the scale dependence is scale-free essentially.

The power law between Moran's *I* and granularity can be explained by the derivation in last Section. The variance of each variable is calculated, then its change with segmentation time $n$ is fitted to exponential function (only for the part when $n$ = 6~13). It is found that the variance of all p variables and most t variables are significantly exponentially correlated with $n$ at the 0.05 level (the corresponding critical value is 0.707, as shown in Table 4). Due to the exponential relationship between $\varepsilon$ and $n$, it can be derived that $V(\varepsilon)$ of most variables follow a power law against $\varepsilon$ significantly at the 0.05 level. According to Eq. 18, $I(\varepsilon)$ follow a power law against $\varepsilon$ as well.

**Table 4** The correlation coefficient of exponential fit to the change of variance with segmentation time $n$ (only for $n$ = 6~13). Variance follow exponential function against $\varepsilon$ significantly except for some t variables that are marked red.

| City | p variable | t variable | City | p variable | t variable |
|---|---|---|---|---|---|
| Beijing | 0.965 | **-0.698** | Nanjing | 0.979 | 0.905 |
| Changsha | 0.987 | 0.961 | Qingdao | 0.982 | 0.829 |
| Chengdu | 0.983 | 0.916 | Shanghai | 0.978 | **0.436** |
| Chongqing | 0.983 | 0.892 | ShenGuan | 0.977 | **-0.457** |
| Dalian | 0.983 | 0.890 | Shenyang | 0.971 | 0.809 |
| GuangFo | 0.967 | **-0.063** | Tianjin | 0.967 | 0.844 |
| Hangzhou | 0.973 | 0.855 | Wuhan | 0.983 | 0.881 |
| Harbin | 0.959 | 0.812 | Xi'an | 0.963 | 0.808 |
| Jinan | 0.960 | 0.836 | Zhengzhou | 0.974 | 0.884 |

| Kunming | 0.982 | 0.905 |

Calculation and analysis so far, we can reach the key part of empirical study in this paper. The mathematical relationship between Moran's *I* and correlation dimension derived above, is to prove theoretically that there is sometimes scaling phenomenon behind spatial autocorrelation. With the calculation results in this section, the validity of the internal relationship between spatial autocorrelation and fractal scaling deduced above can be verified. It can be found from Eq. 18 that

$$I(\varepsilon)V(\varepsilon) \propto \frac{C(\varepsilon)}{\varepsilon^4 N(\varepsilon)} = \varepsilon^{D_0+D_2-4} = \varepsilon^{-\gamma}. \tag{23}$$

For the part when $n = 6\sim13$, calculate the capacity dimension $D_0$ and correlation dimension $D_2$ in each city according to Eqs. 3 and 5 respectively, and the scaling exponent $\gamma$ of variable $I(\varepsilon)V(\varepsilon)$ changing with $\varepsilon$ according to Eq. 23. It is found that the sum of $D_0$, $D_2$ and $\gamma$ is about 4 (Table 5), which verifies the derivation above. In fact, since the limit condition of $N(\varepsilon)\to\infty$ cannot be met, and the power relationship between $V(\varepsilon)$ and $\varepsilon$ does not always hold, $D_0+D_2+\gamma$ is slightly different from 4. The use of t variables can lead to distortions in the original spatial pattern of built-up area, which results in the same $D_0$ values as $D_2$. Additionally, due to its definition, t variables do not conform to Eq. 9, the summed value of $D_0+D_2+\gamma$ is more dissimilar from 4 when compared to using p variables.

**Table 5** The scaling exponent changing with granularity $\varepsilon$ in each city (only for $n = 6\sim13$). The total value of $D_0+D_2+\gamma$ is very close to 4, and the value calculated using p variables is closer to 4 than when using t variables.

| City | Based on numerical variable p | | | | Based on classification variable t | | | |
|---|---|---|---|---|---|---|---|---|
| | $\gamma$ | $D_0$ | $D_2$ | $D_0+D_2+\gamma$ | $\gamma$ | $D_0$ | $D_2$ | $D_0+D_2+\gamma$ |
| Beijing | 0.317 | 1.947 | 1.916 | 4.180 | 0.126 | 1.893 | 1.893 | 3.911 |
| Changsha | 0.234 | 1.877 | 1.804 | 3.915 | 0.121 | 1.745 | 1.745 | 3.611 |
| Chengdu | 0.254 | 1.918 | 1.868 | 4.040 | 0.097 | 1.826 | 1.826 | 3.750 |
| Chongqing | 0.283 | 1.861 | 1.790 | 3.934 | 0.212 | 1.723 | 1.723 | 3.658 |
| Dalian | 0.166 | 1.797 | 1.735 | 3.699 | 0.094 | 1.696 | 1.696 | 3.486 |
| GuangFo | 0.284 | 1.944 | 1.913 | 4.141 | 0.123 | 1.889 | 1.889 | 3.902 |
| Hangzhou | 0.340 | 1.917 | 1.869 | 4.126 | 0.176 | 1.820 | 1.820 | 3.815 |
| Harbin | 0.231 | 1.839 | 1.778 | 3.848 | 0.143 | 1.737 | 1.737 | 3.616 |
| Jinan | 0.232 | 1.873 | 1.824 | 3.929 | 0.144 | 1.786 | 1.786 | 3.715 |
| Kunming | 0.207 | 1.807 | 1.739 | 3.754 | 0.141 | 1.687 | 1.687 | 3.516 |
| Nanjing | 0.290 | 1.884 | 1.820 | 3.994 | 0.181 | 1.761 | 1.761 | 3.703 |
| Qingdao | 0.203 | 1.878 | 1.826 | 3.906 | 0.098 | 1.788 | 1.788 | 3.675 |
| Shanghai | 0.354 | 1.949 | 1.911 | 4.214 | 0.132 | 1.876 | 1.876 | 3.884 |
| ShenGuan | 0.228 | 1.906 | 1.864 | 3.997 | 0.105 | 1.836 | 1.836 | 3.776 |
| Shenyang | 0.226 | 1.882 | 1.831 | 3.939 | 0.113 | 1.794 | 1.794 | 3.701 |

| Tianjin | 0.252 | 1.907 | 1.855 | 4.014 | 0.127 | 1.820 | 1.820 | 3.767 |
| Wuhan | 0.248 | 1.883 | 1.814 | 3.945 | 0.135 | 1.762 | 1.762 | 3.660 |
| Xi'an | 0.268 | 1.912 | 1.869 | 4.049 | 0.115 | 1.834 | 1.834 | 3.783 |
| Zhengzhou | 0.287 | 1.910 | 1.861 | 4.059 | 0.143 | 1.813 | 1.813 | 3.769 |
| avg | 0.258 | 1.889 | 1.836 | 3.983 | 0.133 | 1.794 | 1.794 | 3.721 |

In summary, the scale dependence of Moran's *I* follows power law, but only within a certain scale range, that is, there exists a scaling range. There are two reasons for the gap between calculation results and theoretical derivation. On the one hand, derivation in this paper relies heavily on the limit condition $\varepsilon \to 0$, which is difficult to satisfy in reality. On the other hand, when $\varepsilon$ is too fine, land use class is over-segmented, Moran's *I* is squeezed by the upper limit of 1, and its change with granularity shifts from power law to logarithmic function. While the overall distribution is still scale-free, and still can be approximated by power law.

## 4 Discussion

Scale dependence of urban measures of spatial autocorrelation is universal due to complexity of spatial pattern of built-up area. The purpose of this paper is to explore the scaling properties and laws behind the scale dependence of spatial autocorrelation. Therefore, based on locality of spatial correlation and fractal properties of urban spatial pattern, the model of Moran's *I* depending on scale is derived, and the theoretical derivation are verified with observed data of 19 largest Chinese cities. Based on the above calculation and analysis, the research results can be summarized as follows. First, the scale dependence of Moran's *I* of urban form follows a power law. The power exponent changing with granularity can be linked to correlation dimension through Eq. 23. Second, in empirical study, the scale dependence of Moran's *I* of urban form only follows power law within a certain scale range, that is, there exists a scaling range, with granularity of about 1000~10m (Fig. 5). In empirical analysis of this paper, removing the coarse granularities where Moran's *I* of built-up area change irregularly, the scale dependence of Moran's *I* of most variables obeys power law. Third, the above power law of the scale dependence of Moran's *I* holds for all 19 Chinese cities and both variable types (p, t). In terms of cities, while the spatial pattern of built-up areas may differ across various cities, their Moran's *I* conform to the same power law after the turning points. In terms of variable types, since t variable is equivalent to raster images with the same resolution, Moran's *I* based on remote sensing images with different resolutions aggregated by majority will also follow the same changing regularity. However, since t variable distorts the original pattern to some extent, its changing law is not as significant as that of p variable.

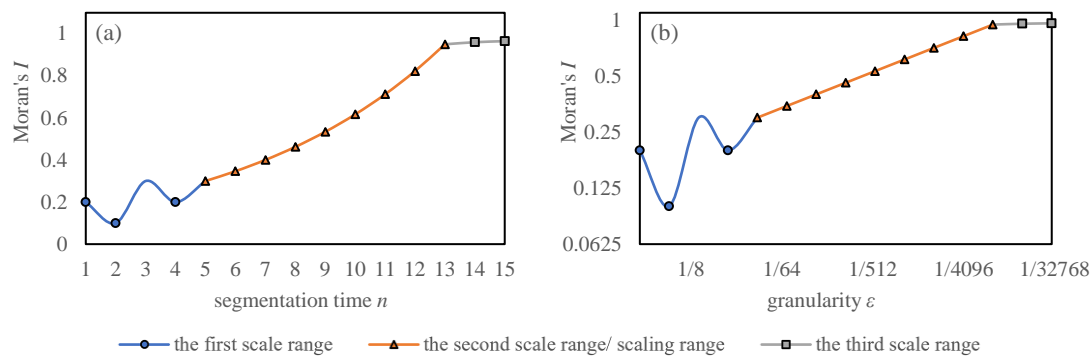

**Fig. 5** Schematic plot of the scale dependence of Moran's *I*. In the first scale range, Moran's *I* fluctuate greatly without obvious regularity; in the second scale range that is actually the scaling range, Moran's *I* follow a power law against granularity; in the third scale range, squeezed by the upper limit 1, Moran's *I* tend to flatten.

Scale-free spatial structure of cities is important for effective explaining urban evolution and proper predicting city development. The research has been ongoing for many years, but a large number of issues have not yet been resolved. In fact, scaling law is an interdisciplinary issue. Scale dependence in geography and ecosystem has attracted the attention of academia for long. Most studies remain in qualitative analysis of scale dependence of Moran's *I*. It is found that at fine granularities, Moran's *I* decrease steadily with the increase of granularity (e.g., Qi & Wu, 1996; Bu et al., 2003; Xu et al., 2004; Tan et al., 2005); however, at coarser granularities, Moran's *I* fluctuate greatly with no obvious regularity (e.g., Overmars et al., 2003; Xie et al., 2006; Qiu et al., 2007; Yin et al., 2008). Therefore, the break point of Moran's *I* changing with granularity is called "intrinsic scale" of spatial autocorrelation in study area (Xu et al., 2007; Chen & Zheng, 2013; Feng et al., 2020; Yuan et al., 2020; Li et al., 2022), and the extreme point of Moran's *I* is called "optimum scale" (Feng et al., 2016; Wang et al., 2021). The concepts of "intrinsic scale" and "optimum scale" are enlightening, but still remain at a qualitative level and are different from "characteristic scale" in mathematical concepts. Chou (1991) quantitatively analyzed the scale dependence of Moran's *I*. Linear, parabolic and logarithmic functions were fitted to the change of Moran's *I* with "resolution level", and logarithmic function was found to have the highest goodness of fit. However, the relationship between Moran's *I* and granularity is not directly established. Zhang et al. (2019) also observed a negative logarithmic relationship between Moran's *I* and resolution, with the characteristic scale of the logarithmic function exhibiting a positive correlation with box dimension. However, these findings have not been theoretically proven, nor have the eigenvalue of the Moran's *I* been given through the characteristic parameters of the logarithmic function. In fact, the characteristic scale of the logarithmic model exceeds the range of Moran's *I*, implying that the scale dependence of Moran's *I* has no characteristic scale, and can instead be approximated by a power law. Moran's *I* is essentially a generalized spatial correlation equation. The author in this paper has derived a spatial autocorrelation function based on variable displacement parameters before, and established the mathematical relationship between spatial autocorrelation function and spatial correlation dimension. Indeed, the relative step function based on variable displacement parameters used to establish the spatial autocorrelation function in that research is similar to the localized contiguity matrix based on variable granularity in this study. Therefore, the results of the two studies can be connected approximately. Through these two different approaches, the link between fractal scaling and spatial autocorrelation has been established.

Compared with previous studies on fractal cities and urban spatial autocorrelation, the novelty of this paper is that it reveals the scale dependence of spatial autocorrelation in urban built-up area from the perspective of fractal scaling invariance. Based on fractal property and locality of spatial correlation, the mathematical expression of Moran's *I* changing with linear scale of spatial partition (that is, granularity) is derived, and the theoretical derivation is empirically analyzed with the help of observation data. According to previous studies and the calculation of this paper, the scale dependence of spatial autocorrelation is divided into two categories: one is the scale dependence with characteristic scales, which can be modeled by functions with characteristic scales (such as exponential function, logarithmic function, Gauss function), and effective spatial autocorrelation measure can be found by model parameters. The other is the scale dependence without characteristic scale, which can only be modeled by power law or its variants, and conventional spatial autocorrelation measure cannot be found in principle (Fig. 6). For the second case, there are two solutions now: one is to extend spatial autocorrelation index to spatial autocorrelation function, and the other is to replace spatial autocorrelation index with scaling exponent (Chen, 2021). Previous studies have found the scale dependence of spatial autocorrelation in urban landscape, or identified a link between fractal pattern and spatial autocorrelation process, but did not distinguish the geospatial properties behind scale dependence based on granularity analysis, nor reveal the specific mathematical expression and physical basis of scale dependence.

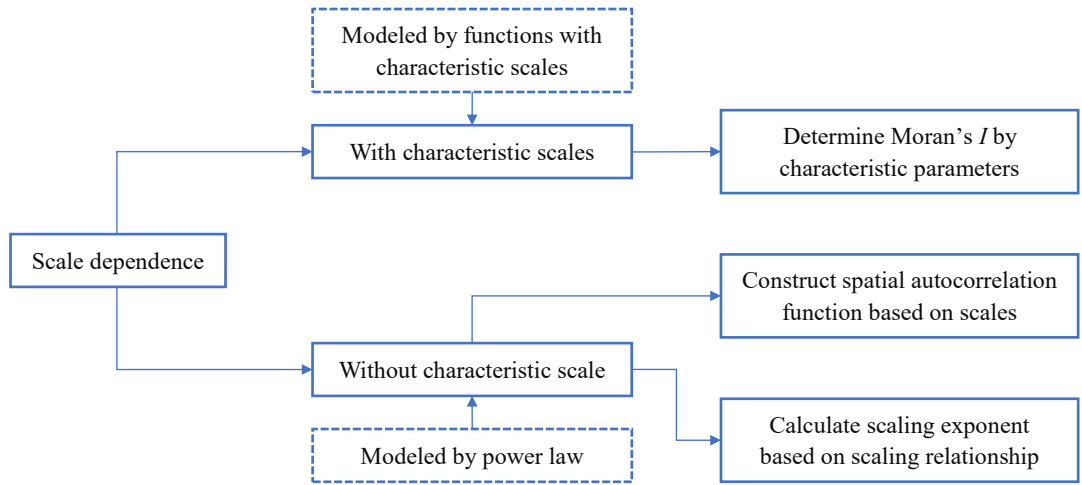

**Fig. 6** Scale dependence of spatial autocorrelation in urban and classification of its solutions. Scale dependence of spatial autocorrelation can be divided into two categories: one is the scale dependence with characteristic scales, which can be modeled by functions with characteristic scales, and effective spatial autocorrelation measure can be found by model parameters. The other is the scale dependence without characteristic scale, which can only be modeled by power law or its variants, and conventional spatial autocorrelation measure cannot be found in principle.

Since empirical research requires lots of work, the scaling analysis for spatial autocorrelation of built-up area in this paper is limited. The main shortcomings of this study are as follows. First, spatial correlation in this work is only based on the postulate of locality, without considering action at a distance. Therefore, the spatial contiguity matrix is constructed based on nominal variables 0 and 1, whose essence is to use step function, which represents local action, to construct spatial weights matrix. Second, the mathematical derivation of this paper is based on the postulates of local spatial correlation and multifractal structure. If fractal properties are undeveloped in urban built-up area, the above derivation are invalid; if fractal properties are developed but the spatial relationship shows significant long-range effects, the results of this paper can only be used approximately. Third, there is no standard method to recognize turning points. The judgement of turning points in function fitting in this study is mainly based on experience and visual inspection, which is somewhat subjective. Indeed, some studies such as Zhang et al. (2019) and Li et al. (2022) have utilized semivariograms to identify the effective scale range of spatial autocorrelation. This approach is more objective and can be applied to recognize turning points in subsequent research.

## 5 Conclusions

The relationship between fractal scaling and Moran's *I* of urban form not only suggests new angle of view of understanding spatial autocorrelation analysis, but also indicates new way of looking at cities. Urban spatial correlation and fractal structure represent different sides of the same coin. Fractal order of cities emerged from bottom-up self-organized evolution by ways of spatial correlation. Based on the above calculation, analysis and discussion, the main conclusions can be drawn as follows. First, the scale dependence of spatial autocorrelation measurement in urban space can be divided into two basic types: with and without characteristic scale. Particularly simple spatial patterns bear no scale dependence. If urban spatial pattern is a relatively simple system, the scale dependence has a characteristic scale, the relationship between Moran's *I* and measurement scale usually follows logarithmic function, and coefficient of the model indicates the characteristic value of Moran's *I*. If urban spatial pattern is a complex system, the scale dependence has no characteristic scale, the relationship between Moran's *I* and measurement scale usually follows power law, and scaling exponent of the model may indicate the characteristic parameter to replace Moran's *I* of urban form. Second, if spatial effect of urban form shows power law, the urban structure follows scaling law. In this case, spatial autocorrelation analysis of cities needs to find another way. Urban patterns bears spatial complexity, and Moran's *I* calculated by conventional method is no longer comparable, even simple spatial autocorrelation index fails. There are two possible ways to solve the problem: one is to extend spatial autocorrelation index to spatial autocorrelation function, which can be used to reveal rich spatial dynamics; the second is to calculate scaling exponent based on the power-law relationship between measurement scale and calculation results, and replace conventional characteristic scale analysis with scaling analysis.

**Acknowledgments**: This research was sponsored by the National Natural Science Foundation of China (Grant No. 42171192). The supports are gratefully acknowledged.